\documentclass[10pt]{article}

\usepackage{amsmath}
\usepackage{amssymb}

\usepackage{times}
\usepackage{enumerate}

\usepackage{graphicx}

\usepackage[labelsep=period,labelfont=bf]{caption}

\usepackage{color} 
\usepackage{colordvi} 

\usepackage[round]{natbib}

\usepackage{bibunits}
\defaultbibliography{okunreply}
\defaultbibliographystyle{hippocampus}

\usepackage{ifthen}

\providecommand{\venue}{}
\newcommand{\arxiv}[1]{\ifthenelse{\equal{\venue}{arxiv}}{#1}{}}
\newcommand{\jns}[1]{\ifthenelse{\equal{\venue}{jns}}{#1}{}}

\renewcommand{\venue}{arxiv}

\usepackage{caption}

\begin{document}
\begin{flushleft}
{\large
\textbf{How (not) to assess the importance of correlations for the matching of spontaneous and evoked activity
\jns{ --  comment on `Population rate dynamics and multineuron firing patterns in sensory cortex' by Okun et al. Journal of Neuroscience 32(48):17108-17119, 2012.}
}}
\\

J\'ozsef Fiser$^1$, M\'at\'e Lengyel$^2$, Cristina Savin$^2$, Gerg\H{o} Orb\'an$^{2,3}$, Pietro Berkes$^4$\\
$^1$Department of Cognitive Science, Central European University, 
$^2$Computational and Biological Learning Lab, Department of Engineering, University of Cambridge, 
$^3$Wigner Research Institute for Physics, Hungarian Academy of Sciences, 
$^4$Volen National Center for Complex Systems, Brandeis University
\end{flushleft}
\vspace{8mm}

\noindent {\large \textbf{Contents}}\\
\begin{tabular}{llr}
February 2013:&Comment on Okun et al.\ (2012)&p.\pageref{part:one}\\
March 2013:&Response to the reply of Okun et al.\ (2013) to our comments&p.\pageref{part:two}
\end{tabular}

\pagebreak

\begin{bibunit}

\noindent {\large \textbf{Comment on \citet{okun12}\label{part:one}}}
\vspace{2mm}

\noindent  \arxiv{In a recent publication, \citet{okun12}}\jns{\citeauthor{okun12}}\ criticise our earlier paper \citep{berkes11} in which we analysed multi-neuron firing patterns in V1 of awake ferrets. We described a remarkably close match between the distribution of multi-neuron firing patterns during spontaneous activity (recorded in the absence of visual stimulation) and during evoked activity recorded while stimulating with naturalistic visual stimuli. We argued that this match confirmed predictions about the cortex implementing a statistically optimal internal model of the environment. In contrast, \citeauthor{okun12}\ claim that this match can be a consequence of trivial statistical properties of multi-neuron firing patterns, specifically fluctuations in overall population firing rates, and as such these epiphenomenal results are not indicative of optimal internal models. Below we explain that the analyses conducted by \citeauthor{okun12}\ suffer from both conceptual and statistical flaws which invalidate their interpretation of their own data. We also show that if the correct analyses are performed on our original data set, the claims of \citeauthor{okun12}\ regarding our findings do not hold and in fact these analyses provide additional support for our main conclusion.

Following the logic of \citeauthor{berkes11}, central to \citeauthor{okun12}'s analysis is the construction of a surrogate data set that respects as constraints some simple statistical properties of the experimentally measured distribution of multi-neuron firing patterns, but is otherwise as random as possible (it has maximum entropy given the constraints). One can then test whether there is a difference between the original results obtained with the real data and the results obtained using surrogate data. If no significant difference can be found, then the original results are deemed epiphenomenal. \citeauthor{berkes11}\ used the simplest possible such surrogate: a distribution under which all neurons fire independently from each other at their true observed firing rates -- this has been a standard test for the role of correlations (of 2nd and higher order) in shaping the distribution of neural responses since the seminal paper of \citet{schneidman06}. \citeauthor{berkes11}\ reported that this surrogate data set did not capture some essential aspects of the real neural responses (\citeauthor{berkes11}, Fig.\  3A-B) and thus that their results could not be trivially explained by single neuron firing rate dynamics.

Building on recent advances in the field of maximum entropy analyses of neural data \citep{tkacik12}, \citeauthor{okun12}\ go one step further and construct a more sophisticated surrogate, which preserves not only single neuron firing rates, but also the distribution of the number of co-active units at a time. This is an important control because, in general, this second constraint introduces correlations at all orders, without any functionally relevant pairwise coupling between neurons. Indeed, the number of degrees of freedom of this surrogate (a measure of its complexity) is still only linear in the number of units, as it is in the simple surrogate that only cares about single neuron firing rates, while in a true second-order surrogate \citep{schneidman06}, that is able to capture the detailed pairwise (functional) connections between neurons, this measure scales quadratically with the number of units. The surrogate used by \citeauthor{okun12}\ is also appealing because it allows a simple biological interpretation: a network in which units are functionally disconnected (or randomly connected) but undergo synchronized fluctuations of activity would produce data whose statistics would match those of this surrogate (see Fig.\  8 in \citeauthor{okun12}). 

\citeauthor{okun12}\ then perform the following key analyses using this surrogate data:
\begin{enumerate}
\item They show that the surrogate data is very close to the true data recorded in A1 (\citeauthor{okun12}, Fig. 3). 
\item They replace both the true spontaneous and evoked activity distributions with their surrogate counterparts and show that some of the results of \citeauthor{berkes11}\ (\citeauthor{berkes11}, Figs.\  2B, 3A-B) can be replicated: the statistical dissimilarity between these surrogates can be minimal in adult animals both in A1 and V1 (\citeauthor{okun12}, Figs.\  4B and 5) and substantially decrease over (simulated) development (\citeauthor{okun12}, Figs.\  7 and 8).
\end{enumerate}

\begin{figure}[!h]
\begin{center}
\includegraphics[width=6.4in]{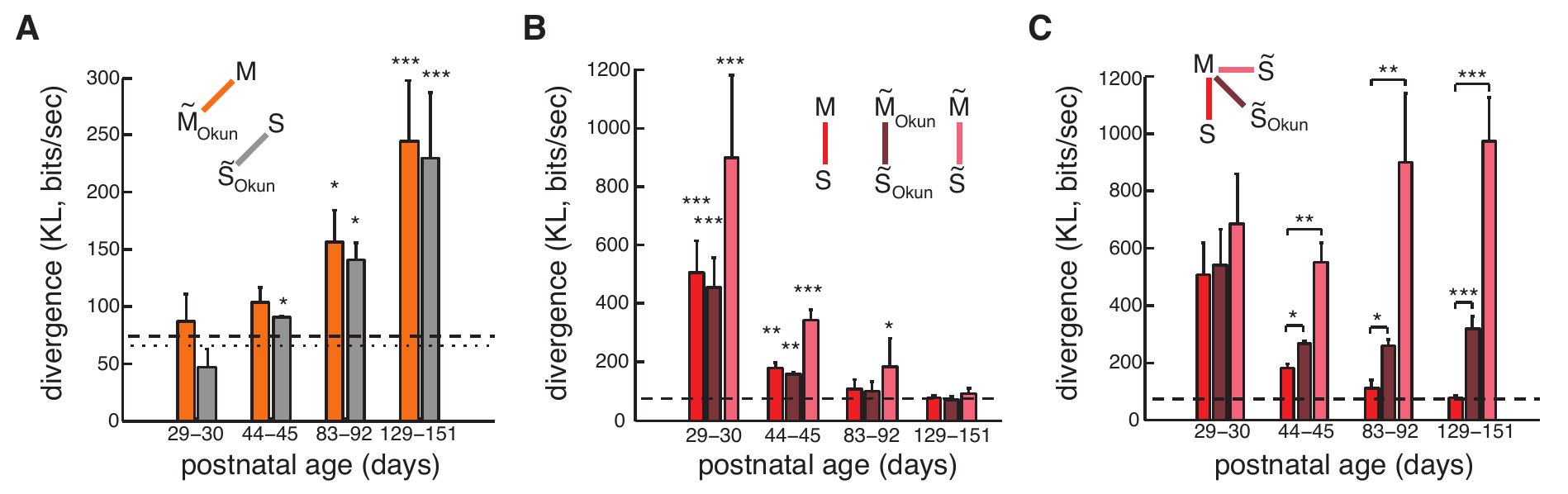}
\end{center}
\caption{~\\ \textbf{A.} Divergences between true activity distributions and their surrogate counterparts. As a reference, the dashed and dotted lines show the average of the within-condition baselines (dotted for S, dashed for M), computed with within-condition data split into two halves. \newline
\textbf{B.} Divergences between movie-evoked (M) and spontaneous (S) activities using the true data (red bars), the surrogate used by \citeauthor{okun12}\ respecting single neuron firing rates and population rate distributions (purple bars), and the original surrogate used by \citeauthor{berkes11}\ respecting only single neuron firing rates (pink bars). Red bars and dashed line are reproduced from \citeauthor{berkes11} \newline
\textbf{C.} Divergences between true movie evoked and the true (red bars) surrogate spontaneous activities (purple bars: \citeauthor{okun12}'s surrogate, pink bars: \citeauthor{berkes11}'s surrogate). Red and pink bars and dashed line (computed as in panel A) are reproduced from \citeauthor{berkes11}\newline
Asterisks denote significant differences from respective baselines (A, B) or between bars (C), $^{*} p < 0.05, ^{**} p < 0.01, ^{***} p < 0.001$, m-test (for details see \citeauthor{berkes11}).
}
\label{fig:old}
\end{figure}

\begin{figure}[!ht]
\begin{center}
\includegraphics[width=6.4in]{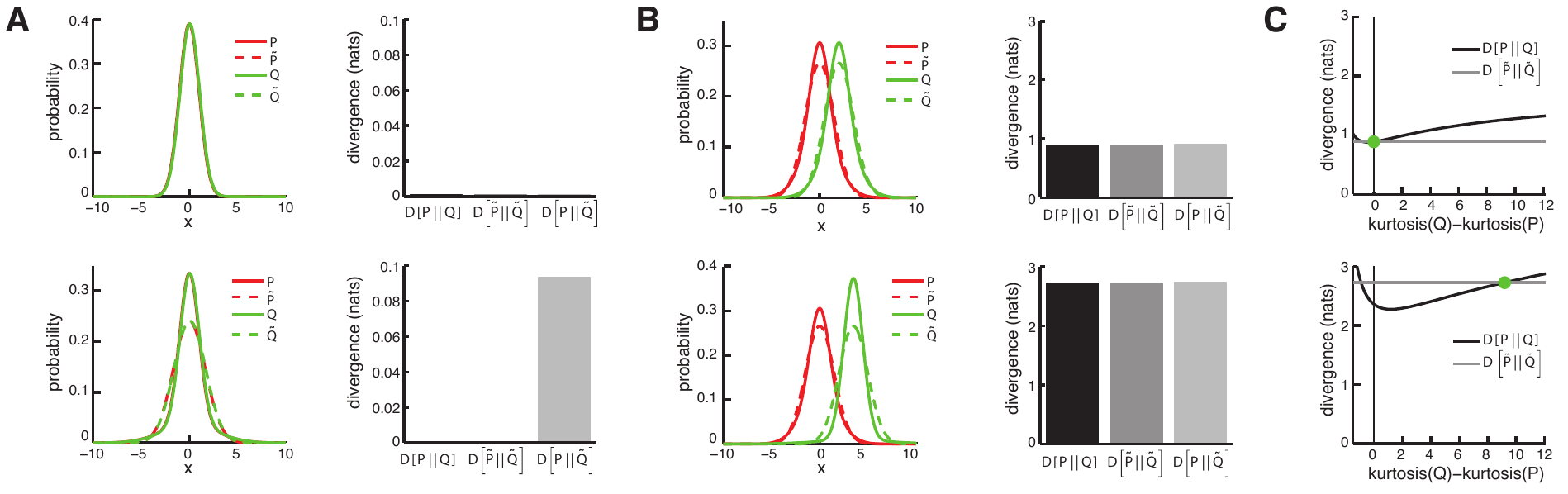}
\end{center}
\caption{~\\ \textbf{A.} When both $\mathrm{D}[\mathrm{P}\|\mathrm{Q}]$ (right panels, black) and $\mathrm{D}[\tilde{\mathrm{P}}\|\tilde{\mathrm{Q}}]$ (right panels, dark grey) are small and they are nearly identical, this does not distinguish between whether the match between $\mathrm{P}$ and $\mathrm{Q}$ is only governed by their lower order statistics (top) or also by their higher order statistics (bottom). Note that using $\mathrm{D}[\mathrm{P}\|\tilde{\mathrm{Q}}]$(right panels, light grey) it is possible to distinguish between these two cases. In the examples shown, $\mathrm{P}$ and $\mathrm{Q}$ (left panels, red and green solid lines) are Gaussian scale mixtures (with two components), and their surrogates ($\tilde{\mathrm{P}}$ and $\tilde{\mathrm{Q}}$, left panels, red and green dotted lines) are moment-matched normal distributions. Panels on the right show respective divergences. True distributions for the top panels are almost exactly Gaussians, hence they are near-identical to their surrogates. True distributions for the bottom panels are leptokurtotic and hence are not identical to their surrogates. $\mathrm{P}$ and $\mathrm{Q}$ are identical in both panels except for a very small shift in their means.\newline
\textbf{B.}  When both $\mathrm{D}[\mathrm{P}\|\mathrm{Q}]$ and $\mathrm{D}[\tilde{\mathrm{P}}\|\tilde{\mathrm{Q}}]$ are large and they are nearly identical, this still does not distinguish between whether the mismatch between $\mathrm{P}$ and $\mathrm{Q}$ is only governed by their lower order statistics (top) or also by their higher order statistics (bottom). Note that $\mathrm{D}[\mathrm{P}\|\tilde{\mathrm{Q}}]$ also does not resolve this issue. Lines in left panels and bars in right panels as in A. For the top panels, $\mathrm{P}$ and $\mathrm{Q}$ differ only in their means. For the bottom panels, $\mathrm{P}$ and $\mathrm{Q}$ also differ in their kurtosis (but not their variance or skewness) which is not controlled by their surrogates. \newline
\textbf{C.} Lines show $\mathrm{D}[\mathrm{P}\|\mathrm{Q}]$ (black) and $\mathrm{D}[\tilde{\mathrm{P}}\|\tilde{\mathrm{Q}}]$ (dark grey) as the kurtosis of $\mathrm{Q}$ is varied relative to that of $\mathrm{P}$, while keeping its mean, variance, and skewness fixed. Bottom and top panels as in B, green points show the kurtoses of the $\mathrm{Q}$ distributions used there. Note that in the case shown in the bottom panel a substantial difference in kurtosis between $\mathrm{P}$ and $\mathrm{Q}$ is accompanied by zero difference in the two divergences. }
\label{fig:new}
\end{figure}

These analyses have questionable relevance for the earlier results of \citeauthor{berkes11}\ on both counts:
\begin{enumerate}
\item Assessing match between surrogate and true data.
\begin{enumerate}[a)]
\item \citeauthor{okun12}\ never demonstrate directly a match between surrogate and true data for their V1 recordings. Given that all the rest of their analyses pertaining directly to the results of \citeauthor{berkes11}\ (obtained in V1) are performed on the V1 data set (\citeauthor{okun12}, Figs.\ 7-8), demonstrating this match would have been a necessary prerequisite for validating their approach.
\item In fact, even for their A1 data, \citeauthor{okun12}'s own analyses do indicate a divergence between the surrogate and true data that appears to be significant (\citeauthor{okun12}, Fig.\  3B, difference between gray and red bars). This significance is never tested statistically, even though if there was a significant divergence that would invalidate their claims about their surrogate reproducing the relevant statistics of the true data (as is the case in their Fig.\  4C, see below in point 2b).
\item Repeating their analysis with their surrogate on the data set used in \citeauthor{berkes11}, we find that the divergence between real and surrogate data significantly increases over development (Spearman's $\rho = 0.68, p = 0.005$ for spontaneous, and Spearman's $\rho = 0.57, p = 0.027$ for movie-evoked activities) and reaches highly significant levels in adult age groups (Fig.~\ref{fig:old}A). 
\end{enumerate}

\item Using surrogate-to-surrogate divergences.
\begin{enumerate}[a)]
\item Measuring the divergence between surrogate spontaneous and surrogate evoked activities, as \citeauthor{okun12} did, is prone to trivial outcomes. Imagine that the original spontaneous and evoked activities are truly being matched by all statistical measures. Then it follows, by definition, that their corresponding surrogate versions, which respect only a subset of these statistical measures, will also be automatically matched to each other.\footnote{As an example, imagine that there are two data sets, each with complicated non-Gaussian looking histograms. We create a Gaussian surrogate for each, matching their mean and variance. If the original histograms looked the same in all their gory detail, their Gaussian surrogates will also look the same -- but this does not mean that the original histograms didn't match in ways that went beyond their mean and variance (Fig.~\ref{fig:new}A). } (This is strictly true only if the divergence between the true distributions is zero, but will usually hold for small but non-zero divergences as well.) As expected, this effect can be demonstrated in the data set of \citeauthor{berkes11} using \citeauthor{okun12}'s surrogate: the divergence between surrogate spontaneous and surrogate movie-evoked activity significantly decreases over development (Spearman's $\rho = -0.75, p = 0.0014$) and reaches non-significant levels in adult animals (Fig.~\ref{fig:old}B, purple bars). To illustrate that this effect is indeed trivial, we computed the same divergence with the simple surrogate used by \citeauthor{berkes11}\ respecting only single-neuron firing rates, and found qualitatively identical results (Spearman's $\rho = -0.71, p = 0.003$, Fig.~\ref{fig:old}B, pink bars). 
\item The appropriate way to establish the importance of higher-order statistics for matching two histograms is to perform analyses of the kind shown in Figure 3B of \citeauthor{berkes11}: surrogate spontaneous must be compared to true evoked activities (or vice versa). If it is only the lower order moments reproduced by the surrogate that matter for the original match between the true distributions, then this divergence using the surrogate should be similar to that computed between the true distributions. By inversion, a substantially larger surrogate-to-true divergence than true-to-true divergence would indicate that higher order correlations were important for the original true-to-true divergence. \citeauthor{okun12}\ show only a single example of such an analysis using their A1 (but not V1) data (\citeauthor{okun12}, Fig. 4C) and even this analysis is inconclusive at best. According to their conclusion from this analysis, not supported by any statistical test, there is no substantial difference between surrogate-to-true and true-to-true divergences -- even though their figure suggests that there is (the green dots are visibly above the diagonal). In fact, \citeauthor{okun12}\ themselves acknowledge that the difference they found in their divergences is due to the inappropriate fit of their surrogate to their real data in the first place (\citeauthor{okun12}, Fig. 3B, see point 1b above). This assessment seems to be correct, but it also points to a crucial flaw undermining their ensuing claims as explained above: when surrogate-to-true divergences are larger than true-to-true divergences then higher-order correlations do play a major role.
\item Most importantly, when the correct surrogate-to-true analysis is performed on our data, we find that in contrast to true-to-true divergences (Fig.~\ref{fig:old}C, red bars), surrogate-to-true divergences do not decrease significantly over time (Spearman's $\rho = -0.32, p = 0.246$) and remain significantly above baseline even in adult animals (Fig.~\ref{fig:old}C, purple bars), further strengthening \citeauthor{berkes11}'s earlier results obtained using the simpler surrogate analysis (Fig.~\ref{fig:old}C, pink bars).
\end{enumerate}
\end{enumerate}

In summary, we concur with the spirit of \citeauthor{okun12}'s work: showing a low (and developmentally decreasing) divergence between spontaneous and evoked activities is necessary but not sufficient for claiming statistical optimality. In fact, this is in line with the approach already taken by \citeauthor{berkes11}\ who provided a set of controls specifically to address this issue, including controls for the effects of spatial correlations, temporal correlations, and stimulus statistics. \citeauthor{okun12}\ elaborate on one of these controls, concerning spatial correlations. Indeed, the surrogate they proposed as an improved control does allow for a more stringent test of trivial factors affecting spatial correlations than the simple surrogate normally used \citep{schneidman06,berkes11}. In that respect, their paper represents an important contribution which will be a useful addition to the toolbox of maximum entropy models in systems neuroscience \citep{schneidman06,tkacik12}. 

Unfortunately, however, there are two reasons why the conclusions \citeauthor{okun12}\ drew from their otherwise useful surrogate are questionable. First, the way they ended up using this surrogate in their analyses was mostly inappropriate to the main question they tried to address, and thus provided trivial results. (And in those cases in which their analysis was appropriate, in their Fig.\  4C, their results seem to agree with those reported in \citeauthor{berkes11}) Second, we performed the correct analyses with their surrogate on our data, and found that these provide additional support for the main conclusion of \citeauthor{berkes11}: the match between spontaneous and movie-evoked activities in the visual cortex is remarkable, cannot be explained by trivial statistical properties and it is, therefore, indicative of statistically optimal internal models in the cortex.

\putbib

\vfill
\noindent\jns{This comment has also been published separately as \citet{fiser13}.}\arxiv{An abbreviated version of this text has also been posted as a Letter to the Editor of the \emph{Journal of Neuroscience} as a reply to \citet{okun12}.}

\end{bibunit}




\begin{bibunit}

\clearpage

\noindent {\large \textbf{Response to the reply of \citet{okun13} to our comments}\label{part:two}}
\vspace{2mm}

\noindent Following our comment (above) on their original paper \citep{okun12}, \citet{okun13} recently published a reply. Our summary for where this discussion stands is the following. There are two general points that \citeauthor{okun12} (2012, 2013) made about how the learning of internal models could and should be tested in experiments, which we agree with in principle, although with some important qualifications. There are also two important specific points about their own results and analyses of data, and the way they interpreted those results, that we disagree with and still find technically flawed.  In our view, these flaws will need to be corrected and some of the main conclusions of \citet{okun12} will need to be revisited in light of these analyses. We detail these points below.
\vspace{5mm}

\noindent\textit{\textbf{Agreement on general points}} \\

\noindent\textbf{Looking at KL divergences (or any other similar measure of divergence) between spontaneous and evoked activities alone cannot be taken as conclusive evidence that an internal model of the environment has been brought about by learning.} We could not agree more and have never in fact said otherwise. In our original paper, we wrote that our findings ``do not address the degree to which statistical adaptation in the cortex is driven by visual experience or by developmental programs'' \citep{berkes11}, and thus interpreted our results as a sign of adaptation (learning-driven or otherwise), not learning \emph{per se}.\footnote{A simple content analysis of \citet{berkes11} shows 13 occurrences of words with a root ``adapt'' and 0 occurrence of words starting with ``learn''.} Thus, \citeauthor{okun12}'s statement that our result alone ``does not indicate that an environmental model has been learned'' echoes our earlier views. However, we now also have results from newer experiments using lid-sutured animals that had no normal vision until the moment of the tests \citep{fiser12,savin13}. These results, to which  \citeauthor{okun13} also refer and which we will present in detail elsewhere, have since provided additional evidence clearly indicating, at least in the oldest age group, that visual experience does have a significant role in the specificity of the match between evoked and spontaneous activities. This is compatible with stimulus specificity only emerging in this age group in control animals (\cite{berkes11}, Fig. 4). Strangely,  \citeauthor{okun13} interpret these results of ours as evidence for specifically non-learning related changes driving adaptation in all but the last age group. However, it is important to note that even in the first three age groups our results do not exclude forms of learning that our manipulation (lid suture) did not affect (such as cross-modal learning, learning from low spatial frequencies, etc), or our data recording and analyses techniques were unable to pick up. Thus, \citeauthor{okun13}'s interpretation is not quite warranted.\\

\noindent\textbf{It is important to control for the effects of populations rate fluctuations in maximum entropy model-based analyses of the role of correlations.} We agree with this statement, too. To make the historical context clear: at the time of the publication of \citet{berkes11}, the ``standard'' way of demonstrating the importance of correlations in maximum entropy analyses was to follow \citet{schneidman06} and compare the true data to the simple surrogate that only controls for single neuron firing rates but not for the population rate distribution. This is precisely the practice we followed in \citet{berkes11}. The contribution of  \citet{okun12} and others \citep{tkacik12} is that they suggested a more stringent test, controlling for the effects of population rate fluctuations, which warrants the revisiting of \emph{all} earlier results that used just the simpler surrogate. We have now revisited our results published in \citet{berkes11}, and confirmed them (see above, Fig.~\ref{fig:old}), and it would be interesting to see which other earlier results (such as \citealp{schneidman06}, and, importantly, the results of  \citealp{okun12} themselves, as we argue below) survive this new, more stringent test. 

\vspace{5mm}
\noindent\textit{\textbf{Disagreement on specific results and interpretations in \citet{okun12}}}\\

\noindent\textbf{Population rate fluctuations account for most of the seemingly correlation-driven effects in the data of \citet{okun12} which therefore implies that they offer an alternative explanation of the results of  \citet{berkes11}.} In our comment on their paper (see above) we stated that this conclusion by \citet{okun12} rests on performing the wrong comparison of divergences that were computed with their otherwise valuable new surrogate. In their reply,  \citet{okun13} dismiss our criticism by stating that ``the point [of their paper,  \citealp{okun12}] was not that $\mathrm{P}$ and $\hat{\mathrm{P}}$\footnote{Where their $\hat{\mathrm{P}}$ is $\tilde{\mathrm{P}}_\mathrm{Okun}$ in our notation, and $\mathrm{P}$ and $\mathrm{Q}$ can stand for movie-evoked (M) or spontaneous (S) activities.} are identical'' and that ``the relationship of $\mathrm{D}[\mathrm{P}\|\hat{\mathrm{Q}}]$ and $\mathrm{D}[\mathrm{P}\|\mathrm{Q}]$ was not central [to their analysis]''. However, as we said before and explain in more detail below in a separate section, this is precisely the shortcoming of their approach that should be mended: analyzing those relationships would have been crucial to support their claims about the (un)importance of true pairwise correlations, and the analysis they chose instead to perform (comparing $\mathrm{D}[\mathrm{P}\|\mathrm{Q}]$ with $\mathrm{D}[\tilde{\mathrm{P}}_\mathrm{Okun}\|\tilde{\mathrm{Q}}_\mathrm{Okun}]$) is not only much more indirect in general than the analyses we suggested and performed on our data (computing $\mathrm{D}[\mathrm{P}\|\tilde{\mathrm{P}}_\mathrm{Okun}]$ and also comparing $\mathrm{D}[\mathrm{P}\|\mathrm{Q}]$ with $\mathrm{D}[\mathrm{P}\|\tilde{\mathrm{Q}}_\mathrm{Okun}]$), but in this case it also leads to wrong conclusions. Importantly, while we have taken seriously the issues they had raised and re-analyzed our data using the surrogate they suggested, thus confirming the crucial (and developmentally increasing) importance of true pairwise (and higher order) correlations for the matching of spontaneous and evoked activities in our data set (see above), Okun et al. did not perform the analyses we pointed out were necessary on their data set. As a result, judging from the analyses they have conducted so far, there is no indication that the important effects of true pairwise correlations shown in our data set are not present in theirs -- although the final word can only be said once they perform the correct analyses on their data set.\\

\noindent\textbf{A random network, such as that shown in \citet{okun12}, could reproduce the key results of  \citet{berkes11}} When making this claim, \citeauthor{okun13} continue to ignore one of the most important controls we already presented in  \citet{berkes11}. In fact, precisely to address this issue, we performed in our original paper \citep{berkes11} additional tests using evoked activities obtained using different kinds of artificial stimulus ensembles (block noise or drifting gratings), and showed that the divergence of these evoked activities from spontaneous activity was significantly larger than the original divergence obtained using natural image movie-evoked activities (see Figure 4 therein). This shows that the cortex is indeed specifically adapted for natural stimuli. Given the importance of this control, it is puzzling why nowhere in \citeauthor{okun12}'s original paper (2012) or in their latest reply (2013) there is any reflection on how their random network fares on it. Therefore, in lack of evidence otherwise, we are led to believe that this control clearly rules out the random-network account of our results proposed by \citeauthor{okun12}\\

\noindent As a final conclusion, before we present our detailed argument about the right comparison of divergences below, we want to stress again that in our view the challenge is now for  \citet{okun12} to correct these technical flaws by using the right divergences and comparisons on the data they recorded, and by performing all necessary tests on their simulations, before they draw conclusions about their own results and those presented in \citet{berkes11}. 
\vspace{4mm}

\noindent\textit{\textbf{Detailed comments on the right comparison of divergences}}
\vspace{2mm}

\noindent In their reply to our comment, \citet{okun13} argue that instead of performing the analyses we suggested and performed on our data (computing $\mathrm{D}[\mathrm{P}\|\tilde{\mathrm{P}}_\mathrm{Okun}]$ and also comparing $\mathrm{D}[\mathrm{P}\|\mathrm{Q}]$ with $\mathrm{D}[\mathrm{P}\|\tilde{\mathrm{Q}}_\mathrm{Okun}]$), the comparison of $\mathrm{D}[\mathrm{P}\|\mathrm{Q}]$ with $\mathrm{D}[\tilde{\mathrm{P}}_\mathrm{Okun}\|\tilde{\mathrm{Q}}_\mathrm{Okun}]$ is the relevant analysis to perform.\footnote{To be precise: for their A1 data they did actually perform the tests we propose, but still drew conclusions from the other test, which we think is misleading, for their V1 data they did not even perform the tests that would have been necessary.} In their eyes, the fact that the difference between these divergences is small supports their claim that ``word distributions primarily reflect changes in population rate dynamic''. This is despite the fact that they agree with us that a finding that both divergences are zero (or close to it) is uninformative as to the importance of correlations (or in general, statistics not controlled by maximum entropy surrogates, Fig. 2A).  In contrast, note that the divergence we suggest should be used, $\mathrm{D}[\mathrm{P}\|\tilde{\mathrm{Q}}_\mathrm{Okun}]$, is able to dissect the role of higher-order statistical structure in causing a match between two distributions in this case (Fig.~\ref{fig:new}A).\\

\noindent Interestingly, \citet{okun13} insist that when the individual divergences, $\mathrm{D}[\mathrm{P}\|\mathrm{Q}]$ and $\mathrm{D}[\tilde{\mathrm{P}}_\mathrm{Okun}\|\tilde{\mathrm{Q}}_\mathrm{Okun}]$, are significantly larger than zero, comparing them is still meaningful. Unfortunately, this is simply not the case: in general, comparing $\mathrm{D}[\mathrm{P}\|\mathrm{Q}]$ with $\mathrm{D}[\tilde{\mathrm{P}}\|\tilde{\mathrm{Q}}]$ (where $\tilde{\mathrm{X}}$ is some maximum entropy surrogate of $\mathrm{X}$) can be very misleading for two reasons. First, because the way differences between different kinds of statistics (e.g. moments) of two distributions jointly determine the total Kullback-Leibler (KL) divergence between two distributions is highly non-linear.  Second, because the KL divergence is not a proper metric.  Fig.~\ref{fig:new}B shows two examples in which both $\mathrm{D}[\mathrm{P}\|\mathrm{Q}]$ and $\mathrm{D}[\tilde{\mathrm{P}}\|\tilde{\mathrm{Q}}]$ are large and are very close to each other. However, in one case (Fig.~\ref{fig:new}B, top) they only differ in their lower order moments (which are being controlled by the maximum entropy surrogates), while in the other case (Fig.~\ref{fig:new}B, bottom) they differ both in these lower order moments and in higher order moments (that are not controlled by the maximum entropy surrogates). In general, it can also be seen (Fig.~\ref{fig:new}C) that $\mathrm{D}[\mathrm{P}\|\mathrm{Q}]$ varies non-monotonically as the differences between $\mathrm{P}$ and $\mathrm{Q}$ in higher-order statistics change, such that any arrangement of $\mathrm{D}[\mathrm{P}\|\mathrm{Q}]$  being greater or smaller than $\mathrm{D}[\tilde{\mathrm{P}}\|\tilde{\mathrm{Q}}]$ is possible. Therefore, it is not at all necessarily true that the difference of these two divergences are zero when their higher-order statistics match. This demonstrates that the fact that $\mathrm{D}[\mathrm{P}\|\mathrm{Q}]$ and $\mathrm{D}[\tilde{\mathrm{P}}\|\tilde{\mathrm{Q}}]$ are large and nearly identical, may be a reflection of ``a property specific to word distributions recorded in the sensory cortex'' \citep{okun13}, but it tells us nothing about the role of higher-order statistical structure in causing a mismatch between $\mathrm{P}$ and $\mathrm{Q}$ -- which is what the original claim of \citet{okun12} was about. Note that our suggested divergence cannot resolve this issue either (Fig.~\ref{fig:new}B), and consequently, we never made any conjectures from the large-divergence case.  As a side note, it would be interesting to see whether there are other divergence-comparisons that may be useful in this case.\\

\noindent In summary, while the comparison \citeauthor{okun12} advocate is unable to resolve, in general, the role of correlations in the case of either a match or a mismatch between evoked and spontaneous activities, the comparison we proposed and used throughout can resolve this role in a match (but not in a mismatch). Thus, while using the correct comparison we have convincingly demonstrated that in our data set, correlations do play a major role in shaping neural activities and causing a match between evoked and spontaneous activities, \citeauthor{okun12}  still need to perform these analyses, especially on their V1 data set. Until these analyses are performed, their claims about the (un)importance of correlations in their data set, including the schematic of the relative distances between distributions they presented (Fig.1 in \citealp{okun13}) remains unjustified and potentially misleading.  In fact, looking at the correct divergences in our data set, the version of their schematic would look rather different in adult animals: $\mathrm{P}$ and $\mathrm{Q}$ would be close ($\mathrm{D}[\mathrm{M}\|\mathrm{S}]~=100$, Fig.~\ref{fig:old}B and C red), so would be $\tilde{\mathrm{P}}_\mathrm{Okun}$ and $\tilde{\mathrm{Q}}$ ($\mathrm{D}[\tilde{\mathrm{M}}_\mathrm{Okun} \| \tilde{\mathrm{S}}_\mathrm{Okun}]~=100$, Fig.~\ref{fig:old}B purple), but the cross-distances would be much larger ($\mathrm{D}[\mathrm{M} \| \tilde{\mathrm{M}}_\mathrm{Okun}]~=200$, Fig.~\ref{fig:old}A orange; $\mathrm{D}[\mathrm{S} \| \tilde{\mathrm{S}}_\mathrm{Okun}]~=200$, Fig.~\ref{fig:old}C purple; $\mathrm{D}[\mathrm{M} || \tilde{\mathrm{S}}_\mathrm{Okun}]~=300$, Fig.~\ref{fig:old}C purple).

\putbib
\end{bibunit}

\end{document}